# OpenVanilla – A Non-Intrusive Plug-In Framework of Text Services

*Tian-Jian Jiang*[1,2]*, Deng-Liu, Kang-min Liu, Weizhong Yang*[3]*,*
*Pek-tiong Tan*[4]*, Mengjuei Hsieh*[5]*, Tsung-hsiang Chang, Wen-Lien Hsu*[1,2]
1. Department of Computer Science, National Tsing Hua University
2. Institute of Information Science, Academia Sinica
3. Department of Theatre, Taipei National University of Arts
4. Department of Physics, National Tsing Hua University
5. School of Information and Computer Sciences, University of California Irvine
128 Academia Road, Section 2, Nankang, Taipei 115, Taiwan
{tmjiang,hsu}@iis.sinica.edu.tw

**ABSTRACT**

Input method (IM) is a *sine qua non* for text entry of many Asian languages, but its potential applications on other languages remain under-explored. This paper proposes a philosophy of input method design by seeing it as a *non-intrusive* plug-in text service framework. Such design allows new functionalities of text processing to be attached onto a running application without any tweaking of code. We also introduce OpenVanilla, a cross-platform framework that is designed with the above-mentioned model in mind. Frameworks like OpenVanilla have shown that an input method can be more than just a text entry tool: it offers a convenient way for developing various text service and language tools.

## INTRODUCTION

Most ideograph-based Asian languages consist of thousands of complex characters. It would be impractical to create a huge keyboard to map every possible character. Modern GUI environments, be it Microsoft Windows, Mac OS X, or X11, all come with build-in tools for transforming multiple composition keystrokes into one single ideograph. These tools are known as input methods or IM for short.

IMs are often categorized into "radical-based" or "phonetics-based" methods. With radical-based input methods, users get a character by typing the composing radicals, whereas with phonetics-based ones users get a character by typing the syllables. If there are homophones, a choice will have to be made, and the proper character is selected and entered.

As most modern GUI environments are designed with standardized IM APIs, applications usually do not have to worry if a user is entering Latin characters or Asian-language texts (and, in the latter case, applications need not to know which kind of input method the user is using). That is to say, text entry is transparent to the applications. Likewise, IMs work through a standardized API and need not to know which applications they are serving. Because of this quality, IMs can be seen as *non-intrusive* plug-ins that adds functionalities to a running application. We will explore the potential of this quality in the present research and show that IMs are not only useful for Asian languages, but also for European, African and even artificial languages.

It should be noted that although alternatives to keyboard exist, none of them - speech recognition, handwriting recognition or optical characters recognition (OCR) - equals keyboard in text entry efficiency and accuracy. As we won't see the end of the keyboard's day anytime sooner, input method will continue to play an important role in our desktop environment.

## HUMAN FACTOR, USABILITY AND UI DESIGN ISSUES

### The Keyboard Layout Problem

The QWERTY keyboard was designed for one single language. Modifications have been applied to the QWERTY keyboard to suit the needs of other European languages such as French QWERTY and German QWERTZ keyboards. This is the origin of the "keyboard layout" of modern GUI environments.

Although the idea of keyboard layout seems to have solved the problem of language-specific text entry well, it has limitations. First, it is impossible to enter more than one language with one single layout. For example, a French keyboard may not be capable of entering Polish text. Second, it is impractical to attach many different keyboards to a single device. To have French and Polish keyboard attached to one laptop computer is unimaginable.

Finally, many devices simply cannot accommodate a large keyboard. For example, mobile phones usually have less than 20 keys. This is where keyboard layouts become unfeasible: there is simply not enough space. Using input methods is the solution that is scalable up to multiple language text entry, and down to limited hardware device.

Input methods like T9 [1] have achieved more balance situation between the trade-off of keystroke numbers and keystroke-character conversion collision rates after some research of human factor and language model. It's valuable to mention Hsu's keyboard layout [2], which maps Chinese bopomofo phonemes to 25 keys by phonetic rules and simi-lar alphabets, as the following figure shows, it is more efficient than any traditional layout, which occupies 40 keys.

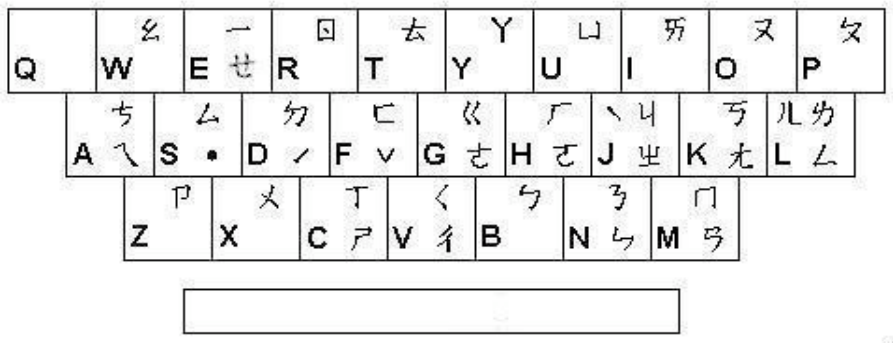

Input method also solves dead-key problems on many keyboards of European languages. When the size of keyboard is limited, using dead keys to input several symbol is also impractical.

### Text Service

So far there are two flavors of definitions of text services. Microsoft describes its Text Service Framework (TSF) by "TSF provides a simple and scalable framework for the delivery of advanced text input and natural language technologies. ... A TSF text service provides multilanguage support and delivers text services such as keyboard processors, handwriting recognition, and speech recognition." On the other hand, Apple defines its Text Service Manager (TSM) as "A text service is a specific text-handling task such as spell-checking, hyphenation, and handling input of complex text." Since this paper is focused on "keyboard" input method, we found it would be interesting to compare with spell-checking task. Although spelling checker is a handy function of MS Word, it is only available there. However, it is possible to develop an input method for this requirement, then a "check as you type" spelling checker will be born.

### Candidate List

During the text entry process, chances are that a sequence of keystrokes maps to multiple characters or phrases. It is especially so for phonetics-based input methods, such as Pinyin for Chinese or Kana for Japanese. A user must then pick up the exact character/phrase he or she wants from a list of possible choices. Such interaction requires displaying those candidate characters/phrases on screen first and waiting for a choice. We call this special UI widget "candidate list".

Candidate list is an indispensable widget for Asian language text entry. However, it can have applications other than picking up a proper character; it could also serve as an on-the-fly spelling checker for European languages. More generally, it is a context-sensitive UI widget for any type of text services.

## HOW INPUT METHOD WORKS

Most modern GUI environments offer a set of low-level API for writing an input method module, but nothing more. A developer that wishes to build up a fully functional IM will have to deal with UI representation and write complex event handlers from scratch. This task has been increasingly complicated on Microsoft Windows and Apple's Mac OS X as their GUI functionalities grow in the recent years. By contrast, the XIM framework has been the de facto standard of IM development for the X-Window environment for years. The last is no easier than its Microsoft or Apple equivalents, though.

Such complexities and difficulties are the reason why many input method frameworks have been flourishing: both IIIMF and SCIM aim to relief the pain of IM development on X Window. The Japanese UIM aims to provide a unified interface to more than two dozens of different IM modules, so that the whole set can be easily ported. OpenVanilla serves Mac OS X and X Window well. In general, a framework should provide a set of abstract API, and takes either a dynamic-loading or client-server approach to work as a mediator between input method modules, operating system, and applications. A framework should also implement a set of default widgets and event handlers. With such "facilities," an IM developer can concentrate on algorithm design with no further concerns on platform-specific details.

Furthermore, when IMs convert source key codes to required target type, they are like output filters more in some sense. Actually, even using the same input method, output results still have chances to be tweaked by applying output filters for different purposes, e.g. conversion between Traditional Chinese and Simplified Chinese.

In short, given that problems of keyboard is still the main input device, human factors of input remain places for improvement, many text processing requirements are still unsatisfied, and input method development are not easy.

## INTRODUCING OPENVANILLA

OpenVanilla is the successor of two successful open source projects on the Mac OS X platform: VanillaInput and SpaceChewing. Both projects were designed to provide input methods that have been (and still are) inadquately supported by Apple's built-in modules. OpenVanilla is designed as an abstract text input/output service framework, and after two major releases (0.6 and 0.7) it currently enjoys a wide user base.

From IM development's point of view, OpenVanilla is designed with the following two principles:

- The framework and a set of various IM modules should be easy to deploy.
- The framework should offer a unified, platform-independent interface to save IM developers' times on investigating complex platform-specific issues. It should allow anyone with some basic knowledge of C/C++ to be able to write his or her own IM module.

OpenVanilla is divided into two parts: platform-dependent loaders and (mostly) platform-independent IM modules. OpenVanilla is actually a very thin layer of interface between the two parts, because it consists of only two C++ header files. This makes OpenVanilla, especially its IM modules, extremely easy to port and deploy.

### What OpenVanilla Is

- It is a set of simple header files
- It is a loose set of Loaders and Modules - but both are tied with the simple and unified set of interface

- It offers a unified interface to every key-event handler (OVKeyCode, OVBuffer, OVCandidate, OVService)
- It is UTF-8 based, using the C-style string definition internally and throughout the framework (plain-old and dirty char* is a good thing, says OpenVanilla)

### What OpenVanilla Is *Not*

- It is not grandiose in design (hence avoiding the over-design syndrome)
- It does not try to be a jack-of-all-trade framework
- It does not claim to solve all IM problems (e.g. hand-writing pad, voice input, etc.) - it aims to solve 95% of the IM problems *elegantly* and is happy with what it can do
- It does not claim to be "cross-platform" without real implementation
- It does not require complicated protocol or deep dependency tree
- It does not care the implementation detail of respective Loader/modules, although it encourages platform-independency under the UNIX tradition (lowest dependency, high fault tolerance, "just work" and "Do What I Say" philosophy, minimum configuration with well-defined default behavior)

### What OpenVanilla Does Not Do:

- No mouse event (would result in 90% code doing 1% of rarely used features)
- No fancy candidate window control (would result in platform-dependency)
- No complicated configure widget design (would require dependency on certain GUI library e.g. gtk).

## SOME CURRENT DEVELOPMENTS

OpenVanilla's simplicity and flexibility makes it easy to *bridge* with other IM frameworks. A bridge module can be written to use other frameworks' IM modules and then loaded by OpenVanilla's Loader. Such is the case of OpenVanilla's UIM bridge that currently offers an Anthy (Japanese Hiragana) input method module.

Another way of bridging is to write a loader in the form of a framework's module. This enables OpenVanilla's modules work on other frameworks. Such is the case of OV's SCIM bridge which makes OpenVanilla successfully work on X Window environment. It's SCIM loader, OVLoader-SCIM, is actually an SCIM-compliant input method module, which in turns loads other OpenVanilla modules dynamically.

Because there was an issue that the candidate list of OpenVanilla 0.6 could not be displayed in Dashboard of Mac OS X 10.4 Tiger, a display server was added into the most recent releases of OpenVanilla. It handles all graphical events, and it makes the design of user interface is separated from back-end control logic, such a design is identical to the spirit of MVC.

Also there is an experimental prototype of a socket-based input method mechanism. Each IM process could be considered as a network client, it requests text services from a server via a TCP socket. Then, text services could be separated from user interface to realize an ideal software engineering structure. It discloses possibilities to combine alternative systems simultaneously, for instance, there could be more choices of programming languages, but not limited by system-level API and the compatibility of Inter-Process Communication. The network connection gives more possibilities for Internet services, too. For example, the back-end information could be obtained via on-line dictionaries or search engines.

## CUSTOMIZATION, SIMULATION AND PERSONALIZATION

Writing an IM even needs no programming training. There is a most important module of OpenVanilla called "Generic Module", it could use a plain text file as an IM, and anyone who knows how to use a text editor could generate his/her own IM easily.

Such a text file is called ".cin" which is given by it's file extension. This kind of data format was invented by the XCIN project, which is another IM framework for X Window system. It uses a two-rows table to record how keyboard events are converted. And, five variables shared by almost all IMs could be configured within such a file format, too. These variables are listed in the following:

- Show candidates as you type.
- Maximum key sequence length.
- Commit at maximum key sequence length.
- Keys are used to choose candidates.
- Use space key to choose the 1st candidate.

The Built-in IMs of MS Windows and Mac OS all use similar approach to customize IM modules, however, the shared configurations are still required. Therefore the ".cin" format has more generality than others. Based on the ".cin" format, developers of OpenVanilla designed a generic IM module and successfully applied it on several popular Chinese IMs

In order to approve the generic IM module, and therefore to make it evaluating to a generic IM engine, the parts of file I/O, data structures and algorithms were replaced by database systems. Based on the advantage of SQLite, developers could use SQL commands but not design an algorithm by themselves to process wildcard searching or sorting, they no longer need to worry about the system performance or reliability while loading or saving data streams.

By editing the ".cin" file or the data in SQLite databases, it is a piece of cake for users to customize the character binding to some keys. Those ones who have specific needs can generate a simplest IM plug-in easily without the help of a programmer. Researchers can generate different IMs very fast by taking the same way when they need experiments or analysis, especially for constructing differed simulating environment while studying IMs for mobile devices.

For personalization, some modern IMs would like to "learn" about users' behaviors and adjust the order of can-didates dynamically. One of our ongoing tasks is taking the concept of Cache Management Pattern [3] as the principle to design such a function. Usages must be tracked and cached first, and then information could be mined from them.

## SHOWCASES

The ".cin" format originated from the famous Xcin project, the first open-source project that offers various Chinese input methods on X11. OpenVanilla's generic IM module uses this open, text-based format to support IMs like Cangjei, Array or Dayi, all of which have their respective user base in the Chinese-speaking world.

Because of OpenVanilla's lean API, a Tibetan IM module that supports four keyboard layouts and character stacking has been written within about 300 lines of codes.

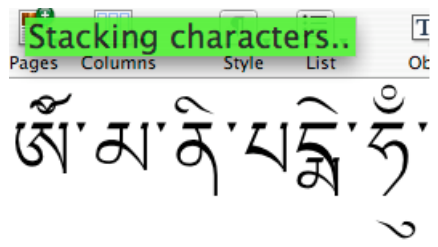

The OpenVanilla Team has also created an IM for Romanized Taiwanese (Peh-oe-ji; POJ). POJ makes extensive use of Latin characters with diacritics. OpenVanilla's POJ module is so far the best and the most flexible among its equals. Because of this module's flexibility, we are confident that it can be adapted to others languages that use Latin script, as many European languages, Vietnamese, or African languages.

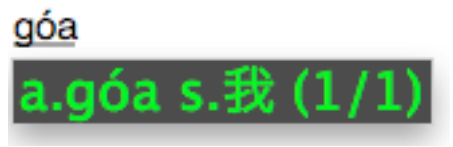

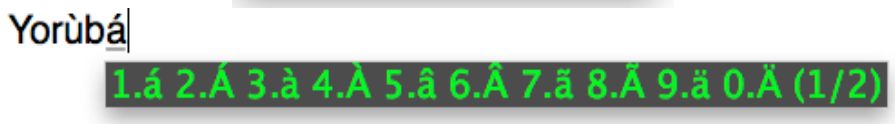

As OpenVanilla is Unicode-compliant, it supports symbol-based input methods with ease. An example is its "EHQ" module, which enables users to type some 1,200+ Unicode symbols, many of them "dingbats," with meaningful mnemonics, e.g. type "*" and you get a long list of available star-shaped dingbats.

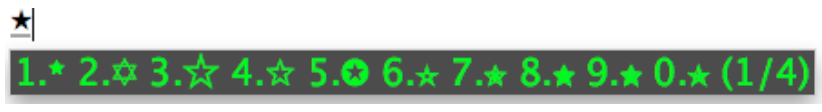

We can even create input methods for European, African, or even artificial or programming languages: such is the case of OpenVanilla's Klingon IM module.

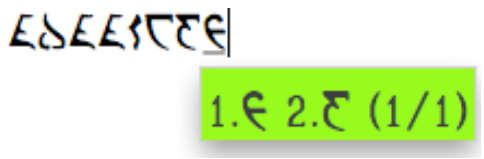

While concerning IMs as output filters, you can use simple codes to convert the form of the results of IMs. OpenVanilla supports Traditional-Simplified Chinese conversion, it is different with other converters that take the strategy to transform user selected sentences, but it converts as you type. It is an outstanding application of output filters. OpenVanilla also supports a module to convert ASCII characters to Morse code, it makes anyone can encode their messages without professional training. OpenVanilla could support a Braille output filter by the same way.

For CJKV users, it's not surprising that intelligent input methods include built-in dictionaries. Actually this approach could be applied on English writing, too. While experimenting socket based input method mechanism, an IM which can notify synonyms in candidate list by using WordNet has been implemented. The user interface migrates from dictionary query application style to input method style.

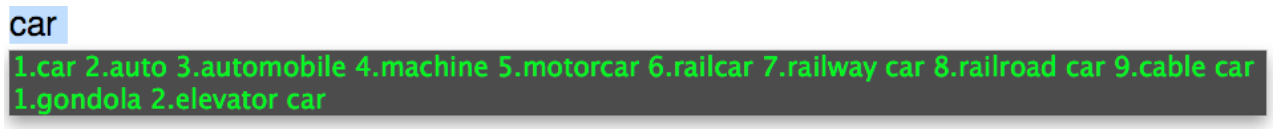

## SUMMARY

IMs should not be limited to only work with Asian languages, but also an alternative of traditional dead key for inputting Latin scripts. Moreover, they are essentials of mobile devices for any language.

Extending to a higher level of text service framework, IM escapes the literal meaning of the word "input". It could be used as spell checking, dictionary, character conversion, and even inline scripting language interpreter. The potential of such text-based application is under-explored.

To make these goals easier to achieve, this paper has proposed a perspective of "non-intrusive plug-in framework of text services", and it has been implemented and proved by OpenVanilla framework. With this scalable framework design and flexible user interface "candidate list", developers and users could release their creativity in a more feasible way to bring potentials to realities.

## FUTURE WORKS

Implementing a bridge between OpenVanilla and Windows IME API is one of the most important tasks in next step, because input method framework is absent on this most popular platform.

The framework design of OpenVanilla is still growing. We wish it to be an experiment platform with positive feedback route between researches and applications. Importing natural language processing techniques [4] to make user-friendly interface is one of the key points in near future.

## ACKNOWLEDGMENTS

The authors would like to acknowledge the contributions of Deng Liu, Kang-min Liu, Weizhong Yang, Pek-tiong Tan, Mengjuei Hsieh and Tsung-hsiang Chang in the developing OpenVanilla.